# Unusually High and Anisotropic Thermal Conductivity in Amorphous Silicon Nanostructures

*Soonshin Kwon[a,\*], Jianlin Zheng[a,\*], Matthew C. Wingert[a], Shuang Cui[a], Renkun Chen[a,b,#]*

[a] Department of Mechanical and Aerospace Engineering, University of California, San Diego, La Jolla, California 92093, United States

[b] Materials Science and Engineering Program, University of California, San Diego, La Jolla, California 92093, United States

[*]These authors contributed equally to this work.
[#]E-mail: rkchen@ucsd.edu

**Abstract:**

Amorphous Si (a-Si) nanostructures are ubiquitous in numerous electronic and optoelectronic devices. Amorphous materials are considered to possess the lower limit to the thermal conductivity ($\kappa$), which is ~1 $W.m^{-1}K^{-1}$ for a-Si. However, recent work suggested that $\kappa$ of micron-thick a-Si films can be greater than 3 $W.m^{-1}K^{-1}$, which is contributed by propagating vibrational modes, referred to as "propagons". However, precise determination of $\kappa$ in a-Si has been elusive. Here, we used novel structures of a-Si nanotubes and suspended a-Si films that enabled precise in-plane thermal conductivity ($\kappa_{\parallel}$) measurement within a wide thickness range of 5 nm to 1.7 μm. We showed unexpectedly high $\kappa_{\parallel}$ in a-Si nanostructures, reaching ~3.0 and 5.3 $W.m^{-1}K^{-1}$ at ~100 nm and 1.7 μm, respectively. Furthermore, the measured $\kappa_{\parallel}$ is significantly higher than the cross-plane $\kappa$ on the same films. This unusually high and anisotropic thermal conductivity in the amorphous Si nanostructure manifests the surprisingly broad propagon mean free path distribution, which is found to range from 10 nm to 10 μm, in the disordered and atomically isotropic structure. This result provides an unambiguous answer to the century-old problem regarding mean free path distribution of propagons and also sheds light on the design and performance of numerous a-Si based electronic and optoelectronic devices.

**Keywords: Thermal conductivity, amorphous silicon, propagon, mean free path, nanostructures, amorphous limit.**

Amorphous Si (a-Si) nanostructures are being broadly used numerous electronic and optoelectronic devices, such as solar cells[1], infrared thermal sensors[2-3], transistors[4-5], and displays[6]. Thermal management of these devices is often critical for their performance, reliability, and lifetime[7]. Thermal transport in a-Si, and more generally in amorphous materials, has been traditionally described by the 'amorphous limit' model that can be traced back to Einstein in 1911[8], who attributed heat conduction in disordered solids to random walk of independent oscillators with a characteristic frequency, known as the Einstein frequency. Subsequently, Slack[9], and Cahill and Pohl[10] refined Einstein's concept and proposed the widely-used minimum thermal conductivity ($\kappa_{min}$) model in disordered solids, referred to as the 'amorphous limit'. This model has worked effectively in explaining $\kappa$ of a large number of amorphous materials, such as oxides[11-13].

However, thermal conductivity of a-Si garnered tremendous renewed interests in recent years as measurements[14-23] showed that thermal conductivity of a-Si can be considerably higher than the amorphous limit for a-Si, which is around ~1 W.m$^{-1}$K$^{-1}$. A summary of prior $\kappa$ measurement results of a-Si is shown in Supplementary Figure S1. While the measured $\kappa$ values scatter, the general trend was that $\kappa$ is close to the $\kappa_{min}$ (~1 W.m$^{-1}$K$^{-1}$) for a-Si, when the film thickness ($t$) is $\leq$ 100 nm. This thermal conductivity is believed to be dominated by non-propagating higher-frequency modes, known as 'diffusons', as originally studied by Allen and Feldman[24-26]. However, for films greater than 1μm thick, $\kappa$ measurement can be higher than 3 W.m$^{-1}$K$^{-1}$. This extra thermal conductivity is believed to be contributed by phonon-like propagating modes, referred to as 'propagons'[24-26].

The observed size dependent thermal conductivity provided insights into the mean free path (MFP) distribution of propagon in a-Si. As the $\kappa$ only starts increasing with the thickness when $t$ is ≥100 nm, these prior results suggested that the lower bound of propagon MFP must be considerably larger than 100 nm, otherwise propagon contribution would have been observed in films with $t$ < 100 nm. This would mean a large discontinuity in the MFP in the transition from diffusons (interatomic distance) to propagons (> 100 nm). However, recent molecular dynamics (MD) and theoretical studies[27-28] showed a smooth transition in the diffusivity, which is proportional to the MFP, from diffusons to propagons. These studies further showed that propagon MFP can range from ~10 nm to ~1 μm, but the bulk $\kappa$ value is considerably lower than 4 W.m$^{-1}$K$^{-1}$. Clearly, there is still no consensus on the MFP distribution of propagons and the bulk '$\kappa$' value of a-Si. Therefore, quantifying the MFP distribution of propagon in a-Si has fundamental significance in understanding thermal management of a large number devices based on a-Si nanostructures. Furthermore, it will shed light on the century-old problem of thermal transport in disordered solids, which is important to the general field of nano-phononics[29-31].

This discrepancy motivated us to re-examine the thermal conductivity of a-Si nanostructures and subsequently quantify its propagon MFP distribution. We realized that all but one[32] of the prior a-Si film $\kappa$ measurements were done along the cross-plane direction. These measurements, most commonly using the 3ω[15, 17] or time-domain thermal reflectance (TDTR) method[16, 18], normally yield total cross-plane thermal resistance, which also includes the contact resistance ($R_c$) between the film and the substrate as well as the metal transducers. For films with very small thickness, e.g., $t$=50 nm, the intrinsic thermal resistance of the films is $R_{a-Si} = t/\kappa \leq 5\times10^{-8}$ m$^2$.K.W$^{-1}$ (assuming $\kappa \geq 1$ W.m$^{-1}$K$^{-1}$), which is on the same order of magnitude as solid-solid interfacial thermal resistance[33]. Thus it is difficult to separate film resistance and $R_c$, resulting in a relative large uncertainty for the measured $\kappa$, especially for thin films ($t < $ 100 nm), as shown in our plot of reported $\kappa$ values (Supplementary Figure S1) and in Braun *et al.*'s work[14]. This issue is further complicated due to the quasi-ballistic transport nature of propagons across the film thickness[34].

In this work, in order to obtain intrinsic thermal conductivity value of a-Si nanostructures without the influence of contact resistance, we utilized novel structures and devices of a-Si nanotubes (NTs) and films that enabled precise in-plane thermal conductivity ($\kappa_\parallel$) measurements over a wide size range of 5 nm to 1.7 μm. The measured $\kappa_\parallel$ showed considerably higher values compared to $\kappa_\perp$: $\kappa_\parallel$ are ~1.5, ~3.0, and ~5.3 W.m$^{-1}$K$^{-1}$ for $t$= ~5 nm, ~ 100 nm, and 1.7 μm, respectively. The size dependent $\kappa_\parallel$ data also suggests that propagons contribute significantly to $\kappa$ of a-Si films even with thickness down to 5 nm, unlike the previously suggested lower bound of 100 nm. We also measured cross-plane thermal conductivity ($\kappa_\perp$) of films, and yielded results that were consistent with prior studies, but considerably lower than $\kappa_\parallel$. The anisotropic $\kappa$ observed in the films further manifests the broad MFP spectra of propagon. With the measured size dependent $\kappa$ along both directions, we extracted the MFP distribution of propagon using an algorithm developed by Minnich[35]. It is found that propagon MFP ranges from 10 nm to over 10 µm and those with MFP greater than 1 µm contributes to ~30 % of $\kappa_p$ in 'bulk' a-Si at 300 K, which has a bulk value approaching ~5.5 W.m$^{-1}$K$^{-1}$ for $t > 2$ µm.

**RESULTS AND DISCUSSIONS**

We measured $\kappa_\parallel$ of a-Si nanotubes with shell thickness ranging from ~5 nm to ~100 nm, and both $\kappa_\parallel$ and $\kappa_\perp$ of a-Si thin films with film thickness ranging from ~26 nm to 1.7 µm.

**Thermal conductivity measurements of a-Si nanotubes and films.** Figure 1 (a) and (b) show the schematic and SEM image of in-plane a-Si NT devices. We also prepared a-Si film samples of 26 nm to 1.7 µm for both $\kappa_\parallel$ and $\kappa_\perp$ measurements. First, we grew a-Si films on either Si or SiO$_2$/Si substrates using identical growth conditions as the a-Si NTs. After the film growth, Figure 1(c) and (d) show the schematic and SEM image of suspended a-Si thin film devices for in-plane and 3ω cross-plane κ measurements (inset in Figure 1c).

**Structural analysis of a-Si nanotubes and films.** Radial distribution function (RDF) analysis (Figure 2e) from the SAED images further confirmed that our a-Si NTs and films have the same degree of atomic disorder as reference a-Si films which were fabricated through ion bombardment on crystalline Si[36].

**Room temperature thermal conductivity of a-Si nanotubes and films.** Figure 3 shows the measured room temperature $\kappa_\parallel$ of a-Si NTs and films as a function of shell or film thickness ($t$). For NTs, we showed that no correlation between $\kappa_\parallel$ and sample length or outer diameter was observed (Supplementary Figure S2), suggesting negligible thermal contact resistance between the NTs and the suspended devices and the similarity between the NTs and films with *t* being the characteristic size. It can be seen that $\kappa_\parallel$ shows strong size dependence, and increases with *t* from 5 nm to 1.7 μm. As discussed before, size dependence was not observed in sub-100 nm a-Si films in prior cross-plane measurements[14, 21]. However, our results showed that $\kappa_\parallel$ start increasing for $t \geq 5$ nm, revealing the important role that propagon plays in thermal transport in a-Si even down to 5 nm thickness. From the size dependent data at 300 K, we can also see that $\kappa$ will saturate to ~5.5 W.m$^{-1}$K$^{-1}$ when the thickness is larger than 2 μm (inset of Figure 3). This 'bulk' value is much larger than what was expected for amorphous solids (~1 W.m$^{-1}$K$^{-1}$).

Figure 3 also shows the thickness dependence of $\kappa_\perp$ at 300 K, which has been corrected after subtracting the contact resistance $R_c$ (Supplementary Note 3). As mentioned earlier, the 3ω method used here can only measure the sum of the intrinsic thermal resistance of the films and $R_c$, which is difficult to separate due to the quasi-ballistic nature of propagon transport across the films and the uncertainty involving in determining $R_c$. Therefore, the $\kappa_\perp$ reported here should only be considered as an effective value and has a larger uncertainty compared to the $\kappa_\parallel$ data, especially for smaller *t*. The measured $\kappa_\perp$ of the Si films are similar to prior $\kappa_\perp$ data (Supplementary Information, Table S1 and Figure S1), namely, only increase for $t \geq 100$ nm, but lower than $\kappa_\parallel$ measured from the films fabricated from the same batches. The observed anisotropy of thermal transport in a-Si films further suggests that the propagon MFP is comparable to film thickness investigated here, similar to the case of crystalline Si films[37-38].

**Temperature dependent $\kappa_\parallel$ of a-Si nanotubes and films.** Figure 4 shows the temperature dependent $\kappa_\parallel$ of a-Si films and NTs from 40K to 300K, along with the 7.7 nm SiO$_2$ NT. The films (*t*=1.7 μm, 525 nm, 170 nm, and 26 nm) and NTs (*t* = ~100, ~40, ~20 and ~5 nm) show similar trend in the temperature dependence. Notably, the ~5 nm a-Si NTs show similar $\kappa$ as that of the SiO$_2$, which is known to possess negligible propagon contribution[28], indicating a similar behavior in a-Si when the size is approaching 5 nm. Furthermore, $\kappa_\parallel$ decreases with lower temperature, and by subtracting the diffuson contribution obtained from previous numerical model by Allen *et al.*[25], we can show that propagon contribution is relatively more pronounced at lower temperature (Supplementary Figure S3).

**Mean free path distribution of propagon in a-Si.** The size and temperature dependence of $\kappa$ in a-Si NTs and films can be utilized to quantify the cumulative MFP distribution of propagons, $F(\Lambda_p)$, at different temperatures. The cumulative MFP distribution for both directions ($F_\parallel(\Lambda_p)$ and $F_\perp(\Lambda_p)$), as shown in Figure 5a, are reconstructed based on the normalized size dependent $\kappa_p$ (i.e., $\kappa_p/\kappa_{p,bulk}$) for both in-plane and cross-plane at 300 K (see details in **Methods**). Since there should be only one MFP distribution function for bulk a-Si, we varied $p$ and $R_c$ values, and found that $F_\parallel(\Lambda_p)$ and $F_\perp(\Lambda_p)$ shows excellent agreement with $p$=0.35 and $R_c = 2\times10^{-8}$ m$^2$.K.W$^{-1}$. Note that the normalized cross-plane data we used to reconstruct $F_\perp(\Lambda_p)$ is after subtracting $R_c = 2\times10^{-8}$ m$^2$.K.W$^{-1}$, and we do not include the 26 nm film due to the large uncertainty (Supplementary Information, Table S1). The value of $R_c = 2\times10^{-8}$ m$^2$.K.W$^{-1}$ is consistent with previous study for Si-SiO$_2$ interface by Lee and Cahill[33]. The best fitting with $p$=0.35, instead of $p$=0, indicates that propagon scattering is partially specular at surface of our a-Si NT and thin film samples. This is not surprising given the fact that the typical wavelength of propagons in a-Si (>2 nm[28]) is much larger than the surface roughness of the a-Si NT (root mean square (rms)=0.815±0.04 nm), as determined from TEM imaging (Supplementary Figure S4).

The extracted MFP spectra show that propagons with MFP down to 10 nm start contributing to κ at 300 K, which is much lower than the lower bound of MFP spectra (>100 nm) suggested by previous experimental studies[14, 21], but is consistent with the recent MD and theoretical predictions[27-28]. In addition, instead of being saturated at 1 μm[21], we found that propagons with MFP greater than 1 μm contribute to ~30% of $\kappa_{p,bulk}$. The role of these long MFP propagons might have been underestimated previously in both experiments[21] (due to the interfacial effect in $k_\perp$ measurements) and simulations[26, 28] (caused by the limited supercell size). The MFP distribution is surprisingly similar to that of phonon in crystalline Si [21, 39-40]. This underscores the long range correlation in the amorphous structure [28].

From the temperature dependent $\kappa_\parallel$ shown in Figure 4, we also reconstructed the $F_\parallel(\Lambda_p)$ at different temperatures, as shown in Figure 5b. The $F_\parallel(\Lambda_p)$ is shifting slightly towards longer MFP with decreasing temperature. The propagon MFP range at 70 K is from 40 nm to 20 μm, longer than that of 10 nm to 10 μm at 300 K. The contribution to $\kappa_{p,bulk}$ from propagons with MFP greater than 1 μm increases from ~30% at 300 K to ~50% at 70 K. It should be noted that here we assumed $p$=0.35 from 300 K to 70 K. However, at lower temperature, propagons with longer wavelength, which are more likely to exhibit specular boundary scattering, are more dominant. Accordingly, the overall $p$ would be higher at lower temperature. Therefore, the extracted propagon MFP shown in Figure 5b could represent the lower bound for T < 300 K.

**Scattering mechanisms of propagons in a-Si.** We used a phenomological model based on the *Landauer* formalism (see details in **Methods**) to understand the scattering mechanism of propagon and its MFP distribution. The size dependent modeling results are shown in Figure 3, along with the experimental data. The modeled in-plane and cross-plane results were fit to the room-temperature size-dependent data along both directions, by using *the same set* of the fitting parameters: *A, B, C, D, E,* and $\eta$, which are adjustable parameters in the scattering terms. The cross-plane data ($\kappa_\perp$) from the 26-nm thick film were not used for the fitting due to the large uncertainty. The corresponding cut-off frequency $\omega_{c,TA}$ for TA and $\omega_{c,LA}$ for LA modes were determined by examining the smooth cross-over of diffusivity for propagon to diffuson (Supplementary Figure S5). The best fitting parameters were found to be: $A = 9.1 \times 10^{-42}$ $s^3 \cdot rads^{-4}$, $B = 4.2 \times 10^{-20}$ $s \cdot rads^{-2} \cdot K^{-1}$, $C = 175$ K, $D = 16.67$ $s \cdot m^{-2} K^{-1}$, $E = 4.4 \times 10^{-3}$ $K^{-2}$, and $\eta = 0.60$ nm. The modeling results show good fitting with *all* the experimental data (again, $\kappa_\perp$ from the 26-nm thick film was not used for the fitting due to the large uncertainty).

For in-plane film modeling result shown in Figure 3, we found that $p = 0$ (fully diffusive boundary scattering) would not fit the $\kappa_\parallel$ data, which is consistent with the conclusion from the MFP reconstruction process (i.e., p = 0.35). Instead, using the '*p*' calculated by the Ziman formula with $\eta$=0.60 nm shows excellent agreement with the $\kappa_\parallel$ data. The best fitted sample surface roughness $\eta$ is only slightly smaller than the experimental measured rms of a-Si NTs (0.815±0.04 nm, see Supplementary Figure S4). This small discrepancy is likely due to the fact that Ziman model tends to under-estimate '$p$'[41]. Furthermore, instead of using Ziman's formula, we found that using a constant $p$ of 0.35 also fits well the in-plane data. This is consistent with the $p$ value obtained from the MFP reconstruction processes from the $\kappa_\parallel$ and $k_\perp$ data at 300 K.

With exactly the same parameters, we also modeled the temperature dependent behavior for NTs and films, as shown in Figure 4. The fitting agrees well with the experimental data for all the films and NTs down to 150 K, while below 150 K, the model shows slightly higher values compared to the experimental results. This indicates that our model for $\kappa_p$ may underestimate the scattering strength for propagons at low temperature, or $\omega_c$ could be different at lower temperature. After fitting with all the temperature data for films and NTs, we also calculated bulk propagon thermal conductivity (i.e., $\kappa_{p,bulk}$ in Equations 1 and 2 in **Methods**), which are 4.9, 4.88, 4.73, 4.48 W.m$^{-1}$.K$^{-1}$ at 300, 200, 100, and 70 K, respectively.

### CONCLUSIONS

We utilized novel structures of a-Si nanotubes and suspended a-Si films to systematically studied size dependent thermal conductivity of a-Si nanostructures. The $\kappa_\parallel$ measurements eliminated the influence of the thermal contact resistance and enabled precise measurement over a wide size range of ~5 nm to 1.7 μm. The $\kappa_\parallel$ data showed unexpectedly high in-plane thermal conductivity ($\kappa_\parallel$ >3 W.m$^{-1}$K$^{-1}$) in a-Si nanotubes and films of ~100 nm thickness, which is further increased to ~5.3 W.m$^{-1}$K$^{-1}$ in 1.7 μm thick film. The measured $\kappa_\parallel$ is significantly higher than that of $\kappa_\perp$ for films of of ~26 nm to 1.7 μm thick. We find that the propagons have MFP spectra ranging from 10 nm to 10 μm at 300 K, and the contribution to propagon thermal conductivity from propagons with MFP greater than 1 μm increase from ~30 % at 300K to ~50 % at 70 K. We also carried out phenomenological modeling to correlate the propagon scattering mechanisms to the observed MFP distribution, and showed that partially specular scattering boundary scattering and the broad MFP spectra account for the large $\kappa$ anisotropy in the a-Si NTs and films.

**METHODS**

**Preparation and thermal conductivity measurement of a-Si nanotubes.**

The a-Si NTs were fabricated by depositing a-Si shells on Ge nanowires at 490 °C with four different nominal shell thicknesses of 5, 20, 40, and 100 nm, and then selectively etching the Ge cores[42]. Detailed geometry information of NTs can be found in Supplementary Information (Table S2). We measured $\kappa_\parallel$ of the NTs using the suspended micro-device method[43-44]. To calibrate our measurement, we measured an amorphous SiO$_2$ NT of similar geometry (shell thickness of 7.7 nm) and large amorphous SiO$_2$ nanowires (diameters of ~250 nm), and obtained results that match well with the established values of a-SiO$_2$ (Supplementary Figure S7).

**Preparation and thermal conductivity measurement of a-Si films.**

We prepared a-Si film samples of 26 nm to 1.7 μm for both $\kappa_\parallel$ and $\kappa_\perp$ measurements. First, we grew a-Si films on either Si or SiO$_2$/Si substrates using identical growth conditions as the a-Si NTs. After the film growth, devices were fabricated for suspended-beam in-plane (Figure 1c,d) and 3ω cross-plane κ measurements (inset in Figure 1c). Details of device fabrication and $\kappa_\parallel$ and $\kappa_\perp$ measurements can be found in Supplementary Information (Supplementary Notes 9, 10, 11).

**Mean free path reconstruction.**

We followed the method proposed by Minnich[35] to reconstruct mean free path distribution from the measured size dependent $\kappa_\parallel$ and $\kappa_\perp$. For both $\kappa_\parallel$ and $\kappa_\perp$, the propagon contribution $\kappa_p$ is obtained by subtracting the diffuson contribution $\kappa_d$[25] from the measured total $\kappa$, i.e., $\kappa_p = \kappa - \kappa_d$. Then $\kappa_p$ is related to the propagon MFP ($\Lambda_p$)[35, 45] via,

$$\kappa_p = \kappa_{p,\text{bulk}} \int_0^\infty S\left(\frac{\Lambda_p}{t}\right) f(\Lambda_p) d\Lambda_p, \quad (1)$$

where $t$ is the film thickness, $\kappa_{p,\text{bulk}}$ is $\kappa_p$ of bulk a-Si, $f(\Lambda_p)$ is differential MFP distribution, and is related to $F(\Lambda_p)$ through $F(\Lambda_p) = \int_0^{\Lambda_p} f(x)dx$; $S\left(\frac{\Lambda_p}{t}\right)$ is the heat flux suppression function, representing the suppression effect on propagon MFP in thin films relative to bulk a-Si. Using integration by parts on Eq. (1), we can obtain[35, 46]

$$\kappa_p = \kappa_{p,\text{bulk}} \int_0^\infty t^{-1} H\left(\frac{\Lambda_p}{t}\right) F(\Lambda_p) d\Lambda_p, \quad (2)$$

where the kernel $H\left(\frac{\Lambda_p}{t}\right)$ is defined as $H\left(\frac{\Lambda_p}{t}\right) = -dS/d\left(\frac{\Lambda_p}{t}\right)$. With the measured $\kappa_p$ for a-Si NTs and films with different thickness $t$ we can reconstruct the smooth cumulative MFP distribution $F(\Lambda_p)$ from Eq. (2) using an algorithm proposed by Minnich[35] based on convex optimization. The suppression function $S\left(\frac{\Lambda_p}{t}\right)$ of films along the in-plane[45] and cross-plane[47] directions are given as

$$S_\parallel\left(\frac{\Lambda_p}{t}\right) = 1 - \frac{3(1-p)}{2}\frac{\Lambda_p}{t}\int_0^1 (u-u^3)\frac{1-\exp\left(-\frac{t}{u\Lambda_p}\right)}{1-p\cdot\exp\left(-\frac{t}{u\Lambda_p}\right)}du, \qquad (3)$$

$$S_\perp\left(\frac{\Lambda_p}{t}\right) = 1 - \frac{\Lambda_p}{t}\left(1 - \exp\left(-\frac{t}{\Lambda_p}\right)\right), \qquad (4)$$

where $p$ in Eq. (3) is the specularity parameter for boundary scattering, which can range from 0 (fully diffusive) to 1 (fully specular). While $p$ depends on frequency and hence MFP, here we employed a single $p$ value in order to reconstruct $F_\parallel(\Lambda_p)$. As we shall see later in the thermal conductivity model, this $p$ value represents an effective one for all the propagon modes when frequency dependent $p$ is taken into account.

**Modeling of thermal conductivity.**

We constructed the analytical model using the *Landauer* approach described by Jeong *et al.*[48] For both in-plane and cross-plane directions, the overall $\kappa$ includes the contribution from propagon and diffuson, i.e., $\kappa = \kappa_p + \kappa_d$. $\kappa_d$ is obtained from Reference[25]. $\kappa_p$ can be modeled based on the *Landauer* approach, namely[48]

$$\kappa_p = \left(\frac{k_B^2 T \pi^2}{3h}\right)\int_0^{\omega_{c,j}}\left(\frac{M_p}{A}\right)(T_p L)W_p d(\hbar\omega), \qquad (17)$$

where $k_B$, $h$, $T$, $A$, $L$ are Boltzmann constant, Planck's constant, temperature, sample cross-section area, and length, respectively; $\omega_{c,j}$ is the cut-off frequency between propagons and diffusons for the $j^{th}$ mode (transverse acoustic (TA) and longitudinal acoustic (LA) modes); $M_p$ is the number of conducting channels at $\hbar\omega$; $W_p$ is the "window" function defined as $W_p = \frac{3}{\pi^2}\left(\frac{\hbar\omega}{k_B T}\right)^2\left(-\frac{\partial n_0}{\partial(\hbar\omega)}\right)$, where $n_0(\omega)$ is Bose-Einstein distribution; and $T_p$ is the transmission function at $\hbar\omega$. It has been shown that the *Landauer* approach is equivalent to that based on the Boltzmann Transport Equations, but provides a convenient approach to model both diffusive and ballistic transport and along different directions[48]. For the films in-plane modeling, we considered both cases by assuming constant specularity parameter '$p$' and by the Ziman formula, i.e., $p = \exp\left(-\frac{16\pi^2\eta^2}{l_p^2}\right)$[41, 49], where $\eta$ is the characteristic dimension of the surface roughness, and $l_p$ is the wavelength of the propagons. The details of the model can be found in Supplementary Information (Supplementary Note 6).

Scattering mechanisms in this model for propagons in bulk a-Si include harmonic Rayleigh-type scattering (with diffusivity expressed as $D_R^{-1} = 3A\omega^4/v_p^2$)[18, 26, 28], anharmonic *Umklapp* scattering ($D_U^{-1} = 3BT\omega^2 \exp(-C/T)/v_p^2$)[14, 25, 28], and anharmonic "two level system" (TLS) scattering ($D_T^{-1} = \frac{D\hbar\omega}{k_B}\tanh\left(\frac{\hbar\omega}{2k_BT}\right) + \frac{D}{2}\frac{ET^3}{1+Ek_BT^3/\hbar\omega}$)[26, 50-51]. Here, '$A$', '$B$', '$C$', '$D$', and '$E$' are adjustable parameters, which were determined by fitting the model to all the measured samples within the studied temperature range. Although the importance of the tunneling states scattering in a-Si is still under debate[32, 52-54] and the tunneling states density for TLS in a-Si strongly depends on the deposition methods[55-56] and growth temperature[54], we found that we had to include the TLS in order to fit the temperature dependent $\kappa_\parallel$ data. If we only considered the Rayleigh-type and *Umklapp* scattering terms, the modeling result would lead to increasing $\kappa$ at low temperature (Supplementary Information, Figure S6).

**Acknowledgments.** This work was supported by National Science Foundation, grant numbers DMR-1508420 (on a-Si films) and CBET-1336428 (on a-Si NTs). We thank Dr. J. Xiang for his help and input on Si NT synthesis.

**Author contributions.** S.K. and J.Z. contributed equally to this work. S.K. synthesized and characterized the materials. S.K. fabricated the devices. J. Z., M.C.W., S.K., and S. Cui performed the thermal measurements. J. Z. carried out the theoretical analysis. R. C.

conceived the experiment. All authors discussed the results, wrote and commented on the manuscript.

**Supporting Information Available:** Additional fabrication, measurement, and modeling details. This material is available free of charge via the Internet at http://pubs.acs.org.

**Competing financial interests:** The authors declare no competing financial interests.

**Figure Captions**

Figure 1. Thermal conductivity measurement schemes for in-plane and cross-plane configurations. (a) Schematic and (b) SEM image of in-plane a-Si NT device. (c) Schematics of either in-plane (main panel) or cross-plane 3$\omega$ (inset) a-Si film devices. (d) SEM image of suspended a-Si film supported by two Pt/Cr electrode bridges (marked as heating and sensing). Scale bars for (b) and (d) are 5 μm.

Figure 2. TEM images of (a) 5 nm thick a-Si NT, (b) 20 nm thick a-Si NT, (c) 100 nm thick a-Si NT, (d) 100 nm thick a-Si film, and (e) Radial distribution function (RDF) of the a-Si NTs and the 100 nm-thick a-Si film. The RDF of a reference a-Si film is from previous work by Laaziri *et al.*[36] The scale bars are 10 nm for (a) and (b), 50 nm for (c), and 5 μm for (d).

Figure 3. Thermal conductivity ($\kappa$) of the a-Si NTs (blue triangle up) and a-Si films for in-plane (red triangle down) and cross-plane (black square). The experimental results of in-plane films agree very well with the a-Si NTs. The cross-plane data have been corrected after subtracting the contact resistance (Supplementary Note 3). The cross-plane 26 nm film shows large error bar mainly due the uncertainty of the contact resistance. Model based on *Landauer* approach[48] is used to calculate the thickness dependent thermal conductivity behavior of thin films. The cross-plane model (black dash line) fits well with our cross-plane data, except the 26-nm thick film which is dominantly affected by contact resistance. The model with specularity parameter p=0 (blue dash-dot line) gives poor fitting with our in-plane thin film and NTs data, while the model with frequency dependent '*p*' by Ziman's formula (roughness $\eta$=0.60 nm) (red solid line) shows excellent agreement with the in-plane data. We also plot the model based on an effective constant *p* of 0.35 for comparison. The inset shows the same plot with linear scale on the x-axis. $\kappa$ is saturating to ~5.5 W/m-K when thickness is larger than 2 μm. Thermal conductivity of the diffuson part (dash-dot-dot line) based on Allen-Feldman (AF) theory[25, 28] is also shown in the inset as a reference.

Figure 4. Temperature dependent $\kappa_\parallel$ for a-Si film and NT samples. Films with thickness of 1.7 μm (gray diamond), 525 nm (black triangle down), 170 nm (blue triangle up) and 26 nm (dark cyan square) show similar trend of temperature dependence as the NTs with thickness of 96 nm (red circle), ~40 nm (dark yellow symbols), ~20 nm (pink symbols) and ~5 nm (dark green symbols). Our model shows excellent agreement with the experimental data for all the samples down to 150 K. At T<150 K, the fitting slightly deviates from the experiment, suggesting the scattering strength for propagons may be underestimated at low temperature. A $SiO_2$ NT with shell thickness of 7.7 nm (cyan diamond) was measured for calibration.

Figure 5. (a) Normalized $\kappa_\parallel$ (red triangle) and $k_\perp$ (black square) of propagons, and the corresponding reconstructed propagon MFP distribution at 300K. The x-axis is the 'sample thickness' for the normalized $\kappa$ or 'MFP$_P$' for the MFP distribution curves. With specularity parameter $p=0.35$ (red solid line), instead of $p=0$ (blue solid line), the MFP distribution reconstructed from $\kappa_\parallel$ agrees well with that from $k_\perp$ (black dash line), suggesting partial specular scattering. The MFP spectra range from 10 nm to 10 μm at 300 K. (b) Reconstructed propagon MFP distributions based on $\kappa_\parallel$ from 300 K to 70 K. The contribution to $\kappa_p$ from propagons with MFP greater than 1 μm increases from 30% at 300 K to 50% at 70 K.

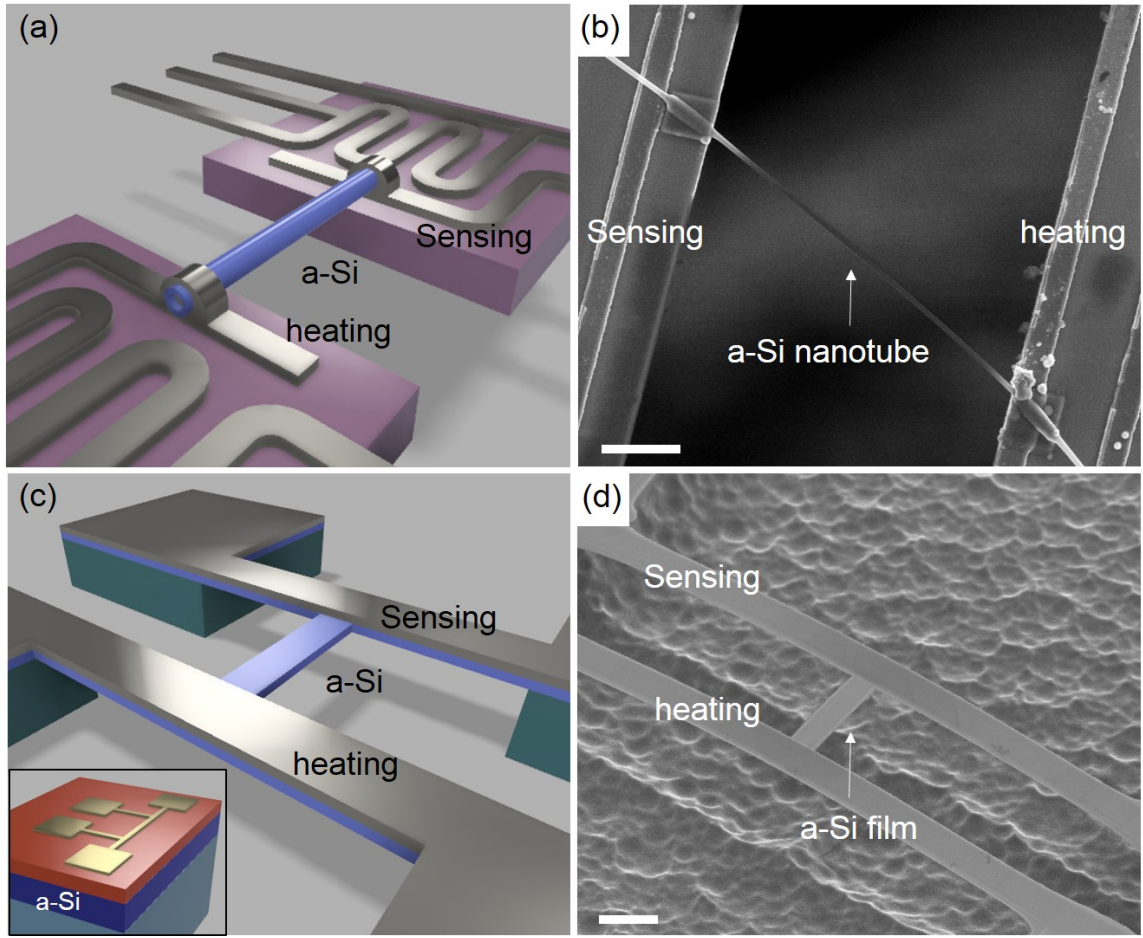

Figure 1. Thermal conductivity measurement schemes for in-plane and cross-plane configurations. (a) Schematic and (b) SEM image of in-plane a-Si NT device. (c) Schematics of either in-plane (main panel) or cross-plane 3$\omega$ (inset) a-Si film devices. (d) SEM image of suspended a-Si film supported by two Pt/Cr electrode bridges (marked as heating and sensing). Scale bars for (b) and (d) are 5 μm.

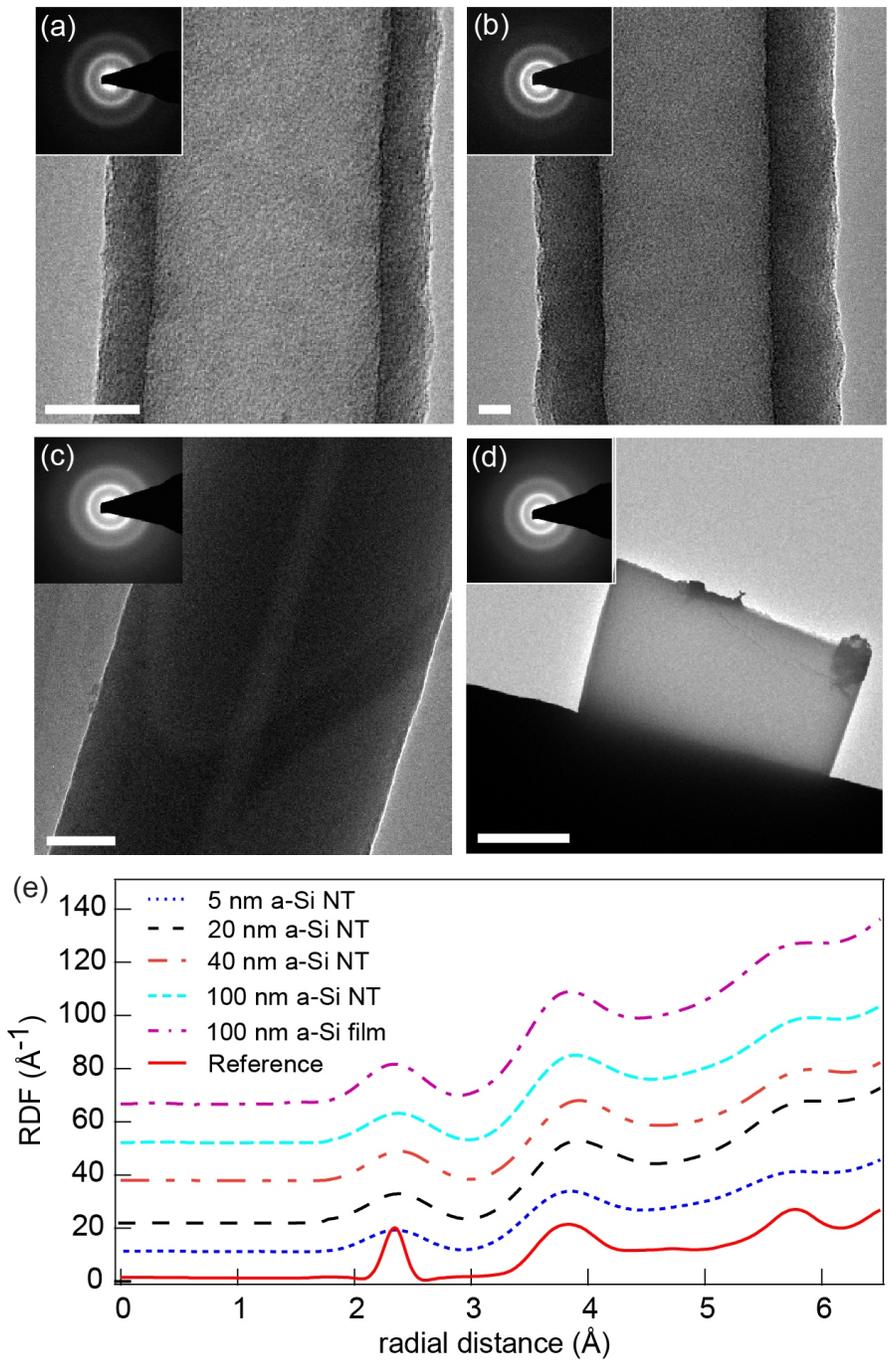

Figure 2. TEM images of (a) 5 nm thick a-Si NT, (b) 20 nm thick a-Si NT, (c) 100 nm thick a-Si NT, (d) 100 nm thick a-Si film, and (e) Radial distribution function (RDF) of the a-Si NTs and the 100 nm-thick a-Si film. The RDF of a reference a-Si film is from previous work by Laaziri et al.[36] The scale bars are 10 nm for (a) and (b), 50 nm for (c), and 5 μm for (d).

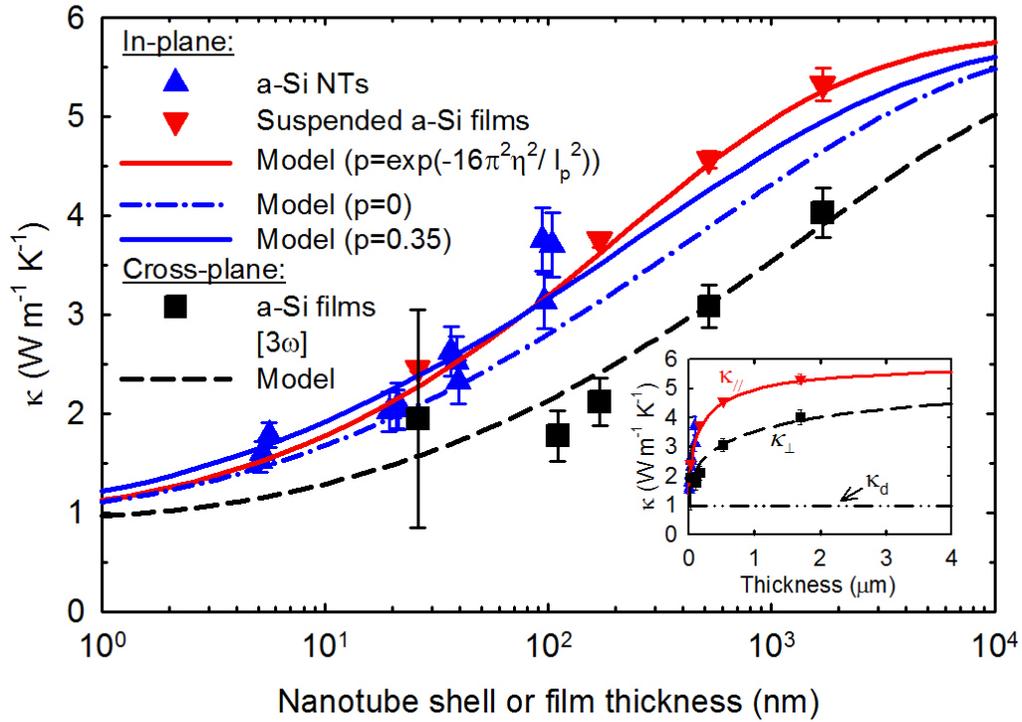

Figure 3. Thermal conductivity ($\kappa$) of the a-Si NTs (blue triangle up) and a-Si films for in-plane (red triangle down) and cross-plane (black square). The experimental results of in-plane films agree very well with the a-Si NTs. The cross-plane data have been corrected after subtracting the contact resistance (Supplementary Note 3). The cross-plane 26 nm film shows large error bar mainly due the uncertainty of the contact resistance. Model based on *Landauer* approach[48] is used to calculate the thickness dependent thermal conductivity behavior of thin films. The cross-plane model (black dash line) fits well with our cross-plane data, except the 26-nm thick film which is dominantly affected by contact resistance. The model with specularity parameter p=0 (blue dash-dot line) gives poor fitting with our in-plane thin film and NTs data, while the model with frequency dependent '*p*' by Ziman's formula (roughness $\eta$=0.60 nm) (red solid line) shows excellent agreement with the in-plane data. We also plot the model based on effective constant '*p*=0.35' (blue solid line) for comparison. The inset shows the same data with linear scale on the x-axis. $\kappa$ is saturating to ~5.5 W/m-K when thickness is larger than 2 μm. Thermal conductivity of the diffusons ($\kappa_d$) (dash-dot-dot line) based on Allen-Feldman (AF) theory[25, 28] is also shown in the inset as a reference.

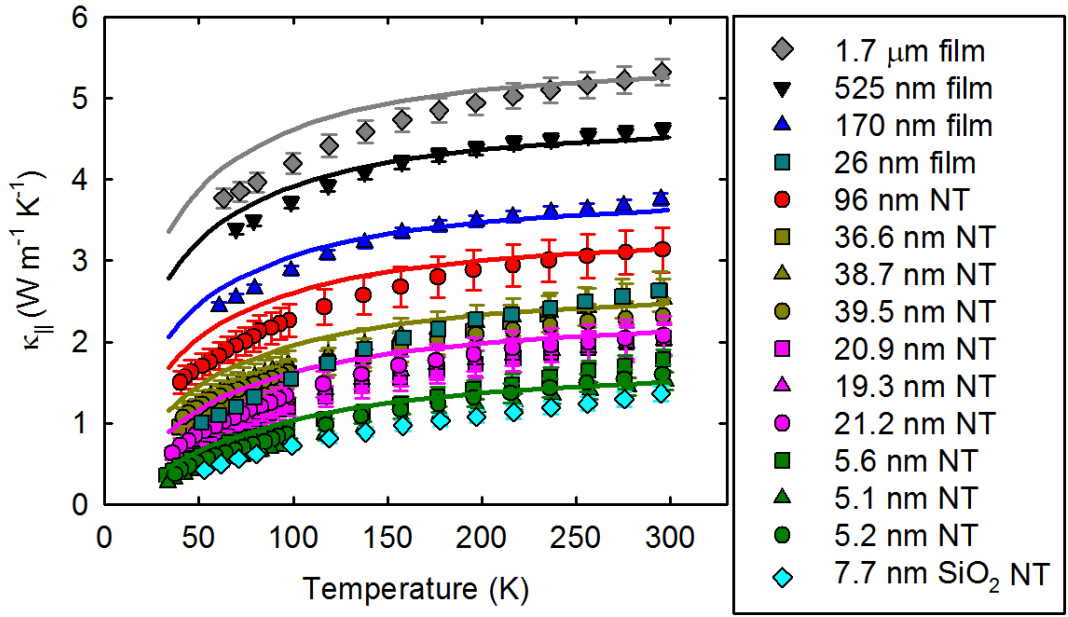

Figure 4. Temperature dependent $\kappa_{\parallel}$ for a-Si film and NT samples. Films with thickness of 1.7 μm (gray diamond), 525 nm (black triangle down), 170 nm (blue triangle up) and 26 nm (dark cyan square) show similar trend of temperature dependence as the NTs with thickness of 96 nm (red circle), ~40 nm (dark yellow symbols), ~20 nm (pink symbols) and ~5 nm (dark green symbols). Our model shows excellent agreement with the experimental data for all the samples down to 150 K. At T<150 K, the fitting slightly deviates from the experiment, suggesting the scattering strength for propagons may be underestimated at low temperature. A $SiO_2$ NT with shell thickness of 7.7 nm (cyan diamond) was measured for calibration.

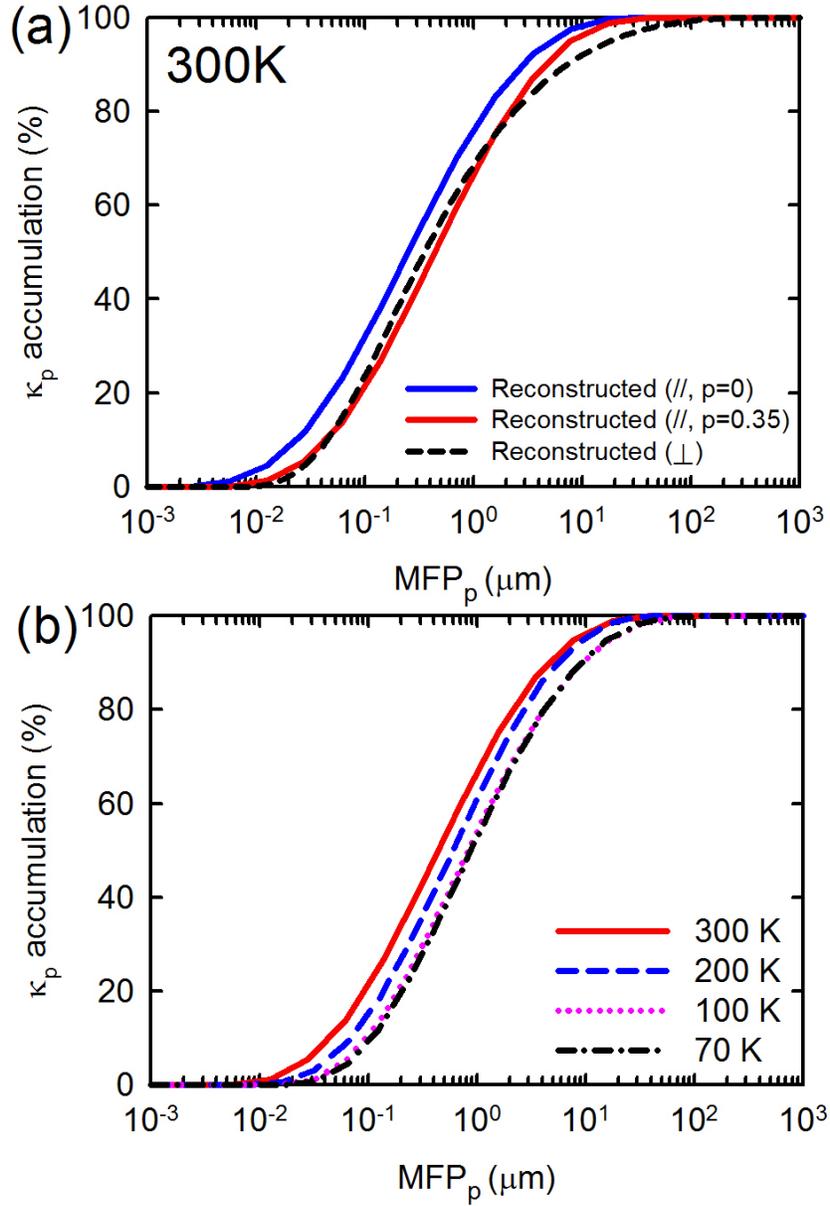

Figure 5. (a) Reconstructed propagon MFP distribution at 300K. With specularity parameter $p=0.35$ (red solid line), instead of $p=0$ (blue solid line), the MFP distribution reconstructed from $\kappa_\parallel$ agrees well with that from $k_\perp$ (black dash line), suggesting partial specular scattering. The MFP spectra range from 10 nm to 10 μm at 300 K. (b) Reconstructed propagon MFP distributions based on $\kappa_\parallel$ from 300 K to 70 K. The contribution to $\kappa_p$ from propagons with MFP greater than 1 μm increases from 30% at 300 K to 50% at 70 K.